\documentclass[twocolumn,showpacs,prl,superscriptaddress,floatfix]{revtex4}
\usepackage{graphicx}
\usepackage{dcolumn}
\usepackage{bm}
\usepackage{amssymb}
\usepackage{amsmath}
\usepackage{amsfonts}
\usepackage{amssymb}
\usepackage{hyperref}
\usepackage{color}
\usepackage{comment}
\usepackage{xcolor}

\begin{document}

\title{Atomic Rydberg Reservoirs for Polar Molecules}
\date{\today}
\author{B. Zhao*}
\affiliation{IQOQI and Institute for Theoretical Physics, University of Innsbruck, 6020
Innsbruck, Austria}
\author{A. Gl\"atzle*}
\affiliation{IQOQI and Institute for Theoretical Physics, University of Innsbruck, 6020
Innsbruck, Austria}
\author{G. Pupillo}
\affiliation{ISIS (UMR 7006) and IPCMS (UMR 7504), Universit\'e de Strasbourg and CNRS,
Strasbourg, France}
\affiliation{IQOQI and Institute for Theoretical Physics, University of Innsbruck, 6020
Innsbruck, Austria}
\author{P. Zoller}
\affiliation{IQOQI and Institute for Theoretical Physics, University of Innsbruck, 6020
Innsbruck, Austria}

\begin{abstract}
We discuss laser dressed dipolar and Van der Waals interactions between atoms and polar molecules, so that a cold atomic gas with laser admixed Rydberg levels acts as a designed reservoir for both elastic and inelastic collisional processes. The elastic scattering channel is characterized by large elastic scattering cross sections and repulsive shields to protect from close encounter collisions. In addition, we discuss a {\em dissipative} (inelastic) collision where a spontaneously emitted photon carries away (kinetic) energy of the collision partners, thus providing a significant energy loss in a single collision. This leads to the scenario of rapid thermalization and cooling of a molecule in the mK down to the $\mu$K regime by cold atoms.
\end{abstract}

\maketitle

\renewcommand{\thefootnote}{\fnsymbol{footnote}} \footnotetext[1]{%
These authors contributed equally to this work.} \renewcommand{\thefootnote}{%
\arabic{footnote}}

There is at present significant interest in preparing and manipulating cold
samples of molecules~\cite{HCN2008,Ye2007,Carr2009}. A promising avenue towards this goal seems to employ
the ubiquitous ultracold atomic gases as cold reservoirs, and to study
mixtures of atomic and molecular gases, where molecules and atoms interact
via collisional processes~\cite{Pillet2006}. Given the well developed tools in manipulating
atoms with external electromagnetic fields~\cite{Metcalf1999}, it is natural to ask whether
we can ``design'' these atom-molecule interactions, thus effectively
engineering an atomic reservoir with desired (collisional) properties. Below
we will describe a specific scenario of engineered elastic and inelastic
collisions involving laser-dressed atoms and ground state molecules with
remarkable, and potentially useful properties. This includes
(i) strong repulsive shields to protect from inelastic collisions and chemical reactions
and (ii) exceedingly large scattering cross sections for
elastic scattering between the atom and the molecule. The relevant energy
(temperature) range includes several mK down to $\mu$K. Equally important,
we will show that (iii) we can design a ``dissipative collision'' where a spontaneously emitted photon carries away (kinetic) energy of the collision partners, thus providing a significant energy loss in a single collision what could be called ``collisional Sisyphus'' effect \cite{Vogl2009}, in analogy to Sisyphus laser cooling of single atoms in external trapping potentials \cite{Dalibard1989}. This suggests rapid thermalization and cooling of a molecule by the cold atom reservoir.

\begin{figure}[tb]
\centering
\includegraphics[width=\columnwidth]{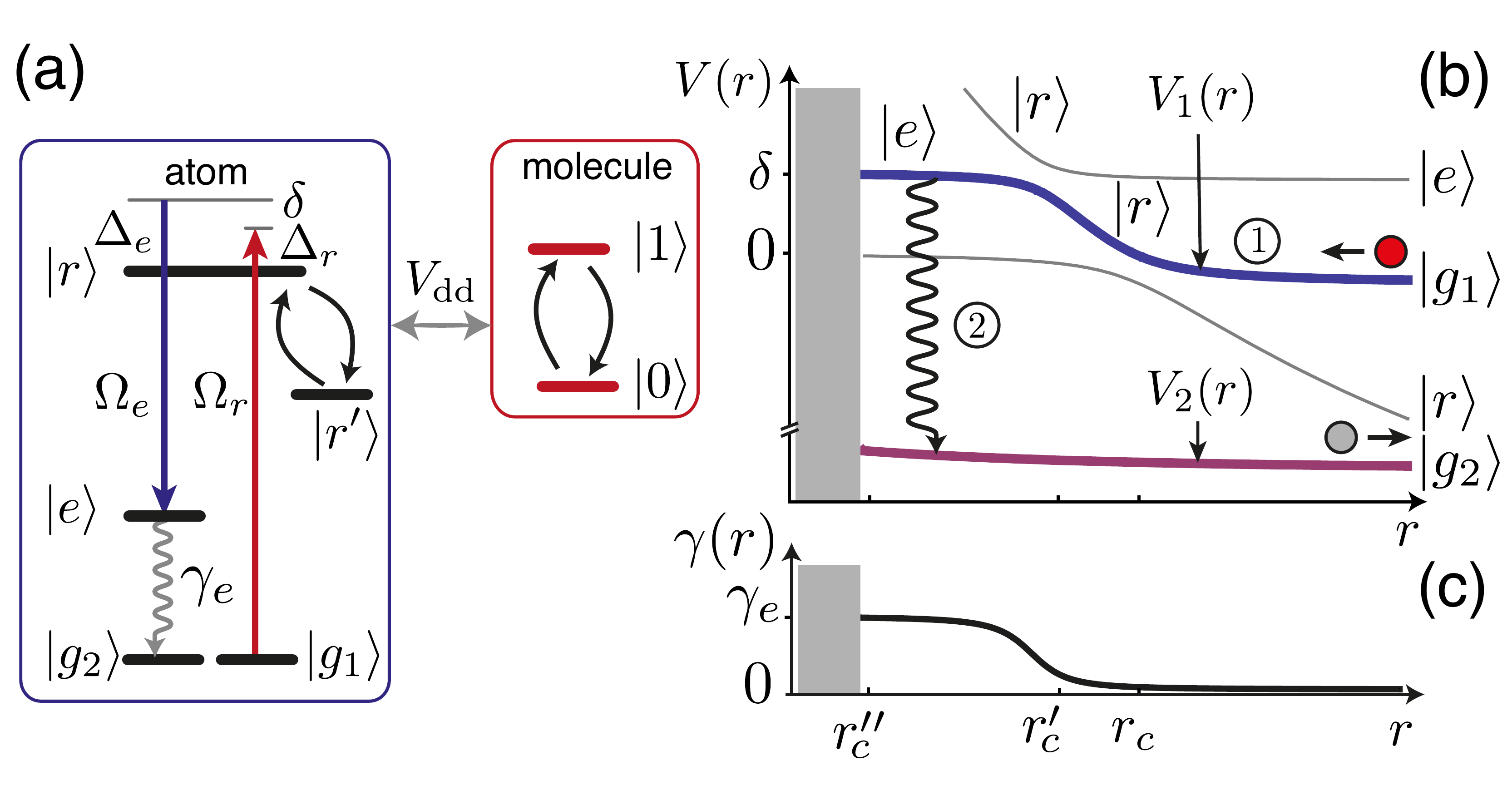}
\caption{
(a) Energy levels of a laser excited atom and a rotational spectrum of a polar molecule. The Rydberg state $|r\rangle$ interacts with the molecule via a dipole-dipole interaction $V_{\rm dd}$ (see text).
(b) Born-Oppenheimer (BO) potentials for the laser dressed atom + molecule complex. We consider a dissipative collision, where (1) the particles collide on the potential curve $V_{1}(r)$ with the atom in $|g_{1}\rangle$, climb the ``blue shield'' step at $r_{c}$, and (2) are quenched to the  potential $V_{2}(r)$ with atom in $g_{2}$. The dominant atomic state is indicated with the molecule in its ground state. (c) Decay rate $\gamma_{1}(r)$ of the BO potential (see text).}
\label{fig1}
\end{figure}

The atomic and molecular level scheme, and the collisional process we have
in mind are illustrated in Figs.~\ref{fig1}(a) and (b), respectively. The basic ingredient is the long range dipolar interaction between molecules in the rovibrational ground state and laser excited Rydberg atoms. The Rydberg state $|r\rangle$ is chosen so that its electric dipole transitions to neighboring states approximately matches the rotational excitation spectrum of the polar molecule with frequencies in the microwave regime (cf. Fig.~\ref{fig1}a), implying a near resonant exchange of molecular and atomic excitations. We focus below on the conceptually simplest configuration, where this interaction reduces to a large {repulsive} and isotropic Van der Waals interaction, $V_{{\rm vdW}}(r)=C_{6}/r^6$, a situation analogous to the large Rydberg-Rydberg interactions underlying the dipole blockade mechanism and the formation of superatoms~ \cite{Gallagher94,Jaksch2000,Saffman2010,HPB2008}. This interaction is admixed to the atomic ground state $|g_{1}\rangle$ with a blue detuned laser $\Delta_r>0$, thus providing an effective interaction between the ground state atoms and molecules. The relevant Born-Oppenheimer (BO) potential for the laser dressed complex is sketched in Fig.~\ref{fig1}(b) as $V_{1}(r)$ (see below), and defines a collision channel for a molecule and atom initially in $|g_{1}\rangle$. In this collision the particles moving adiabatically on $V_{1}(r)$ will encounter a steep ``blue shield'' potential~\cite{Weiner1999} at a distance $V_{{\rm vdW}}(r_c)=\Delta_r$, where typically $r_{c}\gtrsim 100$ nm. By adding a second laser which couples the long-lived Rydberg state $|r\rangle$ down to a low-lying short lived excited state $|e\rangle$ with a detuning $\Delta_{e} = \delta + \Delta_{r} >0$ (see Fig.~\ref{fig1}a), we can add a plateau for $r< r_{c}^{\prime}$ to the adiabatic potential $V_{1}(r)$, so that atoms are efficiently quenched to the ground state $|g_{2}\rangle$ with potential $V_{2}(r)$ according to the rate $\gamma_{1}(r)$ in Fig.~\ref{fig1}(c).

This leads to the following overall picture of collisions illustrated in Fig.~\ref{fig1}(b): (i) for a kinetic energy of relative motion less than the potential step in $V_{1}(r)$, i.e. $E_{\rm kin} < \delta$, we have an elastic collision from an effective hard core potential with (large) radius $r_{c}$; (ii) in a collision with $E_{\rm kin} > \delta$ the particles will climb the potential step  entering the flat dissipative region, which acts as a ``trap door'' so that in a single collision the kinetic energy $\sim\delta$ is carried away by the spontaneous photon with the atom being left in $|g_{2}\rangle$.  Below we will work out a quantitative description of these collisional processes, and argue that they can occur with high fidelity. An essential argument is that during the collision the particles never enter the small distance regime [shaded region in Fig.~\ref{fig1}(b)], and thus the collisional dynamics does not couple significantly to other channels.
The above collision cycle can be repeated by pumping atoms from $|g_{2}\rangle$ back to $|g_{1}\rangle$ so that a significant amount of energy can be lost in a few collisions. Besides, due to the large collision cross sections for elastic processes there is efficient thermalization of the molecules and atoms (sympathetic cooling).

{\em Master equation:} The dynamics of the atom and molecule for a dissipative collision is
described by a master equation, $\dot{\rho}=-i\left[ H,\rho \right] +%
\mathcal{L}\rho $, with the RHS as sum of a Hamiltonian and dissipative
part, and $\rho $ the reduced system density operator after tracing over the
vacuum modes of the radiation field. Such an equation is readily written
down as an extension of the familiar master equations of laser cooling for
atoms by including the molecular dynamics and the atom-molecule
interactions. We neglect, however, recoil kicks from laser absorption
and spontaneous emission, and Doppler shifts, as they provide only small
corrections to our collisional dynamics.

The Hamiltonian has the form $H=\hat{T}+\hat{H}_{I}(\mathbf{r})$, where $\hat{T}=\mathbf{P}^{2}/2M+\mathbf{p}^{2}/2\mu $ is the kinetic energy with $\mathbf{P}$ ($\mathbf{p}$) the
center of mass (relative) momentum and $M$ ($\mu $) the total (reduced)
mass, and $\hat{H}_{I}(\mathbf{r})=H_{0M}+H_{0A}+V_{dd}(\mathbf{r})$ is the
Hamiltonian for the internal degrees of freedom as sum of  a molecular and
atomic Hamiltonian, and the dipole-dipole interaction. For the molecule
we assume a rigid rotor Hamiltonian $H_{0M}=B\mathbf{N}^{2}$ with $B$ the
rotational constant, and $\mathbf{N}$ the angular momentum. For the atomic
Hamiltonian we write in the rotating wave approximation
\begin{equation*}
H_{0A}=\delta \hat{\sigma}_{ee}-\sum_{s=r,r^{\prime }}\Delta _{s}\hat{\sigma}%
_{ss}+\left[ \frac{1}{2}\Omega _{r}\hat{\sigma}_{rg_{1}}+\frac{1}{2}\Omega
_{e}\hat{\sigma}_{re}+\mathrm{h.c.}\right]
\end{equation*}
with notation $\hat{\sigma}_{ij}=|i\rangle \langle j|$ for the atomic
transition operators, and atomic states according to Fig. 1(a). Here we consider the conceptually simplest situation where $|r\rangle=|n,s\rangle$ and $|r'\rangle=|n-1,p\rangle$, with $n$ the principal quantum number and $s$ and $p$ the orbital angular momentum quantum number~\cite{comment0}. By $\Delta _{r}\, (\delta) $ and $\Omega _{r} \,(\Omega _{e})$ we denote the detuning and Rabi frequency of the exciting (quenching) lasers, respectively, and by $-\Delta _{r\prime }$ the energy of the $|r^{\prime }\rangle$ state. For an energy mismatch $E_{0}=E_{a}-E_{m}\ll \min \{E_{a},E_{m}\}$ between the atomic Rydberg states, $E_{a}=E_{r}-E${$_{r\prime }$}, and the rotational splitting of the molecule, $E_{m}=2B$, the interaction is dominated by the dipole-dipole
interactions between the two channels $|r\rangle |0\rangle $ and $|r^{\prime
}\rangle |1\rangle $. The
interaction Hamiltonian between a Rydberg atom and the molecule separated by a distance $
\mathbf{r}=r\hat{\mathbf{r}}$ is $V_{dd}=[\mathbf{d}
_{r}\cdot \mathbf{d}_{m}-3(\mathbf{d}_{r}\cdot \hat{\mathbf{r}})(\mathbf{d}
_{m}\cdot \hat{\mathbf{r}})]/r^{3}(|r\rangle\langle r'|\otimes |0\rangle\langle1|+{\rm h.c.})$ for distances larger than the size of the Rydberg atom, $r>$ $r_{s}\sim n^{2}a_{0}$, where $a_{0}$ is Bohr's
radius. Here, $\mathbf{d}_{r}=\langle r |\hat{\mathbf{d}}|r'\rangle$ and $\mathbf{d}_{m}=\langle 0|\hat{\mathbf{d}}|1\rangle$ are the atomic and molecular transition dipole moments, respectively, with $\hat{\mathbf{d}}$ the dipole operator.
For distances $
r>r_{c_{0}}=(d_{r}d_{m}/E_{0})^{1/3},$ we can adiabatically eliminate $|r^{\prime }\rangle |1\rangle $ and obtain the effective interaction
between ${|r}\rangle $ and $|0\rangle ,$ which is a repulsive and isotropic
van-der-Waals interaction $C_{6}/r^{6}$ \emph{in three-dimensions} with strength $C_{6}=2d_{r}^{2}d_{m}^{2}/3E_{0}.$
Finally, the Liouvillian $\mathcal{L}$ in the master equation describes dissipative
processes due to spontaneous emission. We write $\mathcal{L}=\mathcal{L}_{e}+\mathcal{L}_{p}+%
\mathcal{L}_{b}$, where $\mathcal{L}_{e}=\gamma _{e}\mathcal{D}[\hat{\sigma}%
_{e g_{2}}]$ and $\mathcal{L}_{p}=\gamma _{2}\mathcal{D}[\hat{\sigma}_{g_{2}
g_{1}}]$ account for spontaneous decay from $|e\rangle $ to $|g_{2}\rangle $
and re-pumping from $|g_{2}\rangle $ to $|g_{1}\rangle $, respectively, with
Lindblad term $\mathcal{D}[\hat{\sigma} ]\rho =\hat{\sigma} \rho \hat{\sigma}
^{\dag }-\hat{\sigma} ^{\dag }\hat{\sigma} \rho/2 -\rho\hat{\sigma} ^{\dag }%
\hat{\sigma} /2 $. The third term $\mathcal{L}_{b}\rho$ describes undesired decays,
including in particular spontaneous emission from the Rydberg state, as
discussed below.

{\em Born-Oppenheimer approximation:} We proceed by identifying the BO potentials of the dressed atom-molecule complex as eigenvalues of the internal Hamiltonian $\hat{H}_{I}(r)|i(r) \rangle = V_{i}(r)|i(r) \rangle$ depending parametrically on $r$ [compare Fig.~\ref{fig1}(b)], which, in an adiabatic approximation, provide effective interaction potentials for atoms and molecules.
In particular, the dressed groundstate potential $V_{1}(r)$ corresponds to the BO energy surface that asymptotically connects to the ground state of the atom at large distances, i.e., $|1(r\rightarrow \infty )\rangle \sim |g_{1},0\rangle$. There, atom and molecule are essentially non-interacting. As explained above, the step-like character is obtained in combination with laser dressing on two internal atomic transitions: (i) by coupling $|g_1\rangle$ with $|r\rangle$ in the weak-dressing regime $\Omega_r/\Delta_r < 1$ and for blue detuning $\Delta_r >0$, $V_{1}(r)$ becomes approximately $V_{1}(r) \simeq C_6 / r^6$ for distances $r < r_c$,
with $r\sim r_{c}=(C_{6}/\Delta _{r})^{1/6}$ a resonant Condon point with typical values in the hundreds of nm. This design of interactions is similar to blue-shielding techniques with cold atoms and molecules, however it exploits repulsive vdW-interactions and thus works in three-dimensions. The dominant contribution to the $|1(r)\rangle $ is now $|r,0\rangle $, with $\tau_r \sim n^3$ the lifetime of $|r\rangle$, e.g., in the hundreds of $\mu$s regime for $n\sim 80$~\cite{Saffman2010}.  (ii) A second Condon point can be engineered at distances $r_{c}^{\prime }\equiv (C_{6}/\Delta _{e})^{1/6}<r_{c}$
by weakly admixing $|r\rangle$ with the low-energy excited state $|e\rangle $, using laser light with $\Omega_e/\Delta_e < 1$ and $\Delta_e >\Delta_r$. Here we assume that $|e\rangle $ interacts only weakly with the
molecule and thus $V_{1}(r)$ becomes essentially flat for $r\lesssim r_{c}^{\prime }$. Population in $|e\rangle $ quickly decays to a second groundstate $|g_2\rangle$ at a rate $\gamma_e \sim$ MHz. This makes the decay rate from $|1(r)\rangle$ strongly position-dependent as $\gamma _{1}(r)=\gamma _{e}|\langle 1(r)|e,0\rangle |^{2}$, see sketch in Fig.~\ref{fig1}(c). The BO potential $V_{2}(r)$ with $|2(r)\rangle \sim |g_2, 0\rangle $ is essentially flat,  Fig.~\ref{fig1}(b).

Different BO potentials are coupled via residual non-adiabatic transitions
at $r_{c}$ and $r_{c}^{\prime }$. In particular, population transfer at $r_c$ from $|1(r)\rangle$ to the BO eigenstate that connects to $|r,0\rangle$  for $r\gg r_c$ could induce significant heating and losses. An estimate of the non-adiabatic transition probability can be computed for $b \ll r_c$ within a 1D Landau Zener model as $P_{LZ}= \exp (-2\pi \Omega_r^{2}/(\alpha v))$, with $\alpha $ is the difference of the gradient of the bare potentials at $r_c$. This shows that for any given velocity $v$ (in relative coordinates) non-adiabatic transitions can be always suppressed by increasing $\Omega_r$. Full 3D computations of the non-adiabatic transition probabilities in the semiclassical limit confirm these predictions, see Appendix. Since $|e\rangle$ decays to $|g_2\rangle$, diabatic transitions at $r_c'$ from $|1(r)\rangle$ to the BO eigenstate which adiabatically connects to $|e,0\rangle$ for $r\gg r_c'$ are allowed in our scheme.

Additional diabatic crossings with potential surfaces involving different Rydberg states as well
as attractive resonant dipole-dipole interactions lead to collisional
two-body losses for distances $r\lesssim r_{c_{0}}${. Moreover, interactions between the Rydberg-electron and the molecule play a significant role for $r\lesssim r_s$~\cite{Greene2000}.
These effects are  suppressed by a "blue-shield" at ${r}_{c}^{\prime
\prime }$$>\max\{r_{s},r_{c_{0}}\}$ confining particles'
motion to $r>r_{c}^{\prime \prime }$~\cite{comment1}.

\emph{Reservoir engineering and molecular cooling}: In our scheme, we consider hot molecules undergoing a few scattering processes with cold, interacting, Rydberg-dressed atoms, with lifetime $\tau_d\simeq \tau_r(\Omega_r/2\Delta_r)^{-2}$. Cooling of the molecules comes as a combination of sympathetic cooling with atoms with large elastic cross sections $\sigma \sim \pi r_c^2$ as well as photon-assisted controlled inelastic interactions in a timescale $\tau_c \lesssim \tau_d$ to avoid spontaneous emission from $|r\rangle$ and collisional losses with Rydberg-excited atoms \cite{Alex2011}.

The basic scheme of photon-assisted inelastic collisions can be understood for just an atom and a molecule initially interacting via the BO potential $V_{1}(r)$. It comprises two steps: Firstly, spontaneous emission from $|e\rangle$ couples $|1(r)\rangle $ and $|2(r)\rangle $, according to the spatially-dependent rate $\gamma _{1}(r)$, removing an amount of energy $\lesssim \delta$; secondly, a weak re-pumping laser can transfer population back from $|2(r)\rangle $ to $|1(r)\rangle $, thus closing the cooling cycle. By focussing on $V_{1}(r)$ and $V_{2}(r)$ only and neglecting for a moment unwanted effects described by $\mathcal{L}_{b}\rho$, within the secular approximation the dissipative collisional dynamics in the relative-coordinate frame can be described semiclassically by two coupled Liouville-equations
\begin{equation}
\left( \frac{\partial }{\partial t}+\frac{\mathbf{p}}{\mu }\frac{\partial }{%
\partial \mathbf{r}}\right) f_{i}=\left[ \frac{\partial V_{i}}{\partial
\mathbf{r}}\frac{\partial }{\partial \mathbf{p}}-\gamma _{i}(r)%
\right] f_{i}+\gamma _{j}(r)f_{j}.  \label{eq:LEQ}
\end{equation}%
Here $f_{i}(\mathbf{r},\mathbf{p},t)$ is the Liouville density accounting
for the phase-space distribution of the atom-molecule system in state $%
|i(r)\rangle $ $(i,j\in \{1,2\},i\neq j)$.
\begin{figure}[t]
\centering
\includegraphics[width=0.9\columnwidth]{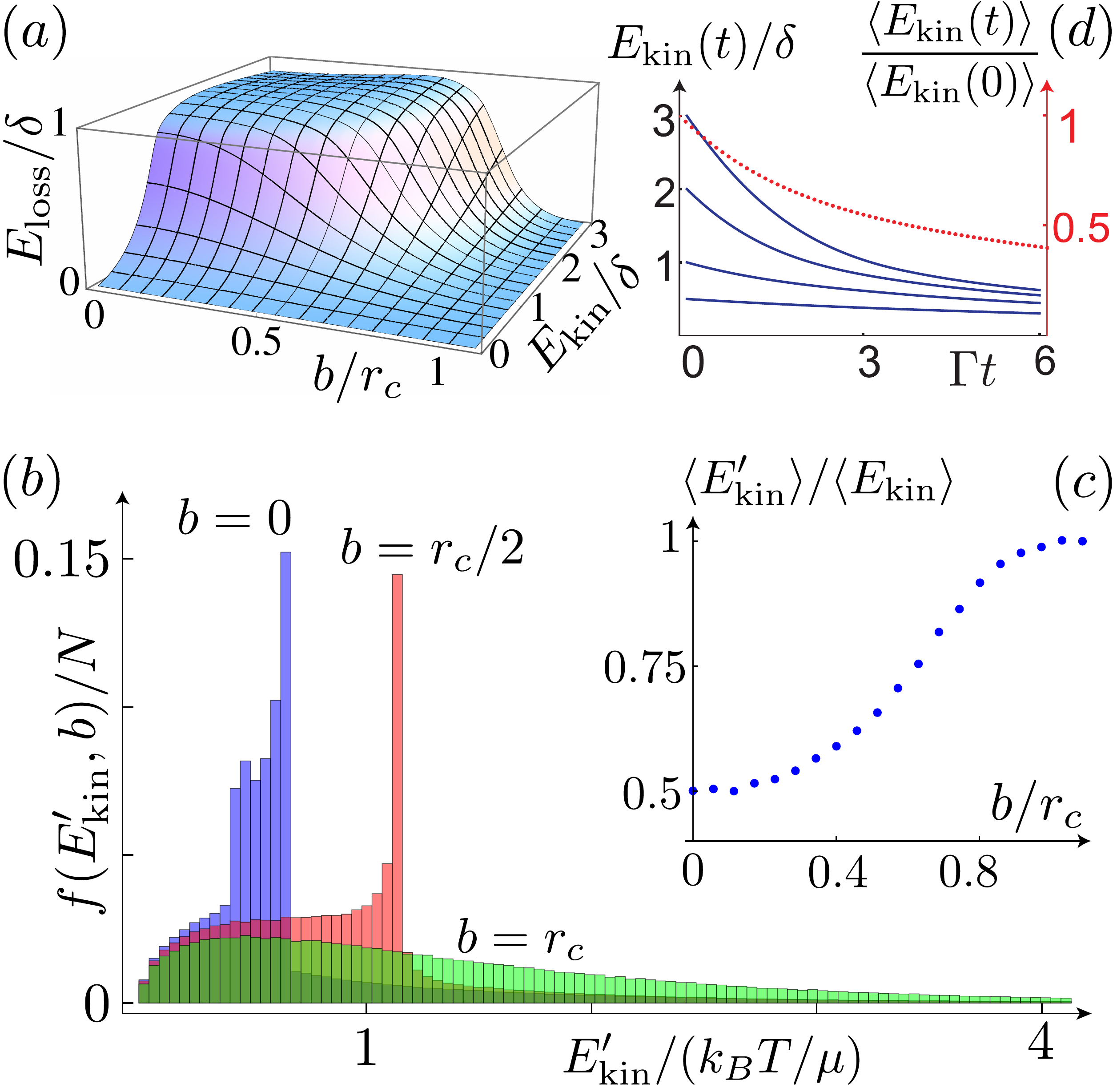}
\caption{Dissipative collisions: (a) Energy loss $E_{\rm loss}$ per collision vs impact parameter $b$ and initial relative kinetic energy $E_{\rm kin}$. (b) Distribution $f$ of
final relative kinetic energies $E'_{\rm kin}$ after a single collision for different $b$, for an initial Boltzmann distribution with $k_{B}T/\delta\approx 0.5$. (c) Average final kinetic energy $\langle E'_{\rm kin} \rangle$ of the corresponding distribution after a single collision. (d) $E_{\rm kin}$ (solid lines, left axis) and the average relative kinetic energy $\langle E_{\rm kin} \rangle$ (dotted line, right axis) vs time with $\Gamma=\rho\sigma v_\mu\approx 2\pi \times 2.3$ kHz the collision rate (see text). Parameters: $d_{m}=7$ Debye, $d_{r}=4400 $ Debye, $E_{0}=2\pi \times2.5$ GHz, $ \Delta _{r}=2\pi \times 60$
MHz, $\Omega _{r}/\Delta _{r}=0.42$, $\delta/\Delta _{r} = 0.33$, $\Omega
_{e}/\Delta _{r}=0.25$, $r_{c}\approx 218$ nm,  $r_{c}^{\prime }/r_{c}\approx 0.95$ , and $r_{c}^{\prime \prime }/r_{c}\approx 0.81$. }
\label{fig2}
\end{figure}
The first term in the RHS is the interaction force, proportional to the gradient of the BO potentials discussed above.
The second and third terms are the spatially dependent decay rate $\gamma _{1}(r)$ and the re-pumping rate $\gamma _{2}$, coupling the two equations. The step-like shape of $\gamma _{1}(r)$ sketched in Fig.~\ref{fig1}(c) reflects the one of $V_{1}(r)$, such that $\gamma _{1}(r) \simeq \gamma_{e}$ for $r_{c}^{\prime \prime }<r<r_{c}^{\prime }$ and $\gamma _{1}(r) \simeq 0$ otherwise. As a result, for incoming relative kinetic energies $E_{\rm kin} < \delta$, particles are reflected elastically at $r \approx r_c$, while
particles with $E_{\rm kin} > \delta$ can reach the region $r < r_c'$ with a velocity  $v^{\prime}=\sqrt{2(E_{\mathrm{kin}}-\delta )/\mu}$ and undergo photon-assisted inelastic collisions. For any given $v'$, spontaneous emission from $|e \rangle$ to $|g_2 \rangle$ [and thus population transfer from $|1(r)\rangle $ to $|2(r)\rangle $] can be made to occur {\it deterministically} in a region of length $d =r_{c}^{\prime }-r_{c}^{\prime \prime }$, by choosing $d$ such that $\gamma _{e}d/v^{\prime } >  1$. This removes an energy of order $E_{\rm loss} \simeq \delta$ in {\it every} single collision, with $\delta$ as large as mK, as shown below. We ensure that the inverse re-pumping from $|2(r)\rangle $ to $|1(r)\rangle$ takes place for distances $r > r_c$ by requiring $\gamma _{2}d/v^{\prime } \ll  1$. In addition, we choose $\gamma _{e}[\Omega _{r}\Omega _{e}/(4\Delta _{r}\delta )]^{2}<\gamma _{2}$ to ensure that the effective Raman transfer rate of population from $|g_{1}\rangle $ to $|g_{2}\rangle $ via $|r\rangle $ is small, and thus the atomic population is in $|g_{1}\rangle $ at $r\gg r_{c}$.
These requirements can be satisfied for realistic atom/molecule configurations, as shown below.

We investigate numerically this dissipative scheme by performing molecular dynamics simulations of the collision of an atom and a molecule, based on Eq.~\eqref{eq:LEQ}. The extension to the case of several atoms and molecules is straightforward. In a semiclassical approximation, the mean energy loss $E_{\rm{loss}}$ in a collision in 3D is computed as $E_{\rm{loss}}=\int [V_{1}(\mathbf{r})-V_{2}(\mathbf{r})]\gamma _{1}(\mathbf{r}_{\rm{cl}})p(\mathbf{r}_{\mathrm{cl}})dt$, where $\mathbf{r}_{\rm{cl}}$ denotes the classical trajectory of the atom/molecule collision, and $p(\mathbf{r}_{\mathrm{cl}})$ is the probability that the atom decays at a given position $\mathbf{r}_{\mathrm{%
cl}}$, with $\dot{p}(t)=-\gamma _{1}(\mathbf{r}_{\mathrm{cl}})p(t)$. As an example, in the calculations we consider a NaH molecule ($d_{m}\approx 7$ Debye) and a Cs atom with Rydberg states $|r\rangle = |46s\rangle$ and $|r'\rangle =|45,p\rangle$, respectively, with $d_{r}\approx 4400$ Debye and $E_{0}=2\pi\times 2.5$ GHz. The laser parameters are $\Delta _{r}=2\pi \times 60$
MHz, $\Omega _{r}/\Delta _{r}=0.42$, $\delta/\Delta _{r} = 0.33$, $\Omega
_{e}/\Delta _{r}=0.25$, so that $r_{c}\approx 218$ nm$,$ $r_{c}^{\prime }\approx $ $208$ nm, and $r_{c}^{\prime \prime }\approx 177$ nm~\cite{comment2}. Figure~\ref{fig2}(a) shows the computed $E_{\rm{loss}}$ as a function of the initial relative kinetic energy $E_{\rm{kin}}$ and the impact parameter $b$. For $b > r_c$ the collision is essentially elastic, as expected. However, for $b \lesssim r_{c}$ and energies $E_{\rm{kin}}>\delta $ the molecule is able to climb the potential step $\delta$ of $V_{1}$ at $r\sim r_{c}$, thus undergoing
deterministic decay to $|2(r)\rangle $.

The effects of this dissipative collisional cooling on a molecular gas with an initial thermal distribution with average temperature $T =0.5$ mK where $\delta\approx 2k_{B}T$ with $k_{B}$ Boltzmann's constant is shown in panel (b). For each fixed value of $b$ we perform $N\simeq10^5$ computations of the collision dynamics, by randomly generating a sample of initial kinetic energies $E_{\rm{kin}}$, according to a Boltzmann distribution. The figure shows the final population distribution $f$ as a function of the final energy $E_{\rm{kin}}'=E_{\rm{kin}}-E_{\rm{loss}}$, for fixed values of $b$, with laser parameters as in Fig.~\ref{fig2}(a). We find that for impact parameters $b>r_{c}$ the distribution $f$ in relative coordinates is largely unaffected by the collision (case $b = r_c$ in the figure). For $b<r_{c}$, however, {\it all population} with initial energy $E_{\rm{kin}}>\delta$ is shifted by an amount $\sim \delta$ towards lower energies. The corresponding average final $E_{\rm{kin}}'$ is shown in Fig.~\ref{fig2}(c) as a function of $b$. For heads-on collisions with $b=0$ approximately $50\;\%$ of the initial kinetic energy is removed after a single collision.

For given $\delta$, $E_{\rm kin}$ and atoms at rest, the energy loss rate is estimated as $- dE_{\rm kin}/dt = \rho (2 E_{\rm kin}/\mu)^{1/2} \mathcal{F}(E_{\rm kin})$, with $\mathcal{F}(E_{\rm kin}) = \int_0^{r_c} E_{\rm loss}(b, E_{\rm kin})2 \pi b db$ and $\rho$ the atomic gas density. The latter is limited by atom-atom interactions of the form $V_{\rm aa} \simeq (\Omega _{r}/(2\Delta _{r}))^{4}V_{rr}$ for atomic distances $r>\rho^{-1/3}_{\rm \max}$, with $V_{rr}=\tilde C_{6}/r^{6}$ the vdW interaction between Rydberg states and $\rho^{-1/3}_{\rm \max}=[\tilde C_{6}/(2\Delta
_{r})]^{1/6}$ a resonant Condon radius ($\rho^{-1/3}_{\rm \max} \simeq 1.7$ $\mu$m for the parameters above). Figure~\ref{fig2}(d) shows the presence of two cooling timescales (solid lines): For $E_{\rm kin}>\delta$, cooling of an energy $\sim \delta$ is achieved on a fast timescale of a few $\Gamma t$, with $\Gamma =\rho
\sigma v_{\mu }\approx 2\pi \times 2.3$ kHz the collision rate, for $\rho =(2$ $\mu $m$)^{-3}$, $r_{c}\approx 218$ nm,  $T=0.5$ mK, $v_{\mu }=\sqrt{3k_{B}T/\mu }%
\approx 0.78$ m/s. For $E_{\rm kin}<\delta$ cooling proceeds slowly, in accordance with the small $\gamma_1(r)$, for $r > r_c$. The same qualitative behavior is found in the average kinetic energy (dots). The lifetime of the dressed state is here $\tau _{d}\approx 1$ ms ($\tau _{r}\approx 45$ $\mu$s), and thus for the parameters above more than $10$ collisions are allowed while cooling. We note that $\delta$ can be dynamically reduced in experiments.

%The interaction $V_{\rm aa}$ decreases quickly with $r$ because of the $r^{-6}$-scaling, and is of the order of $200$ kHz, and thus negligible compared with atom-molecule interactions, already at 2 $\mu $m.

In the lab frame, a molecule loses its energy
due to a combination of both collisional dissipative cooling and sympathetic cooling. The dominant effect depends on the mass ratio $m_{A}/m_{M}$. For an atom initially at rest, an analytic estimate for the atomic and molecular
velocities $v_{A}^{\prime }$ and $v_{M}^{\prime }$ after the collision can be obtained from a simplified model where the total energy is reduced by $\delta $ whenever $\mu v_{M}^{2}/2>\delta ,$ with $v_{M}$ the initial molecular velocity, and is conserved otherwise, as
\begin{eqnarray}
v_{M}^{\prime } &=&V(1-m_{A}/m_{M}\sqrt{1-2\delta /(\mu v_{M}^{2})})  \notag
\\
v_{A}^{\prime } &=&V(1+\sqrt{1-2\delta /(\mu v_{M}^{2})})  \label{eq:vend}.
\end{eqnarray}%
Here, $V=m_{M}v_{M}/M$ is the center of mass velocity. Figure~\ref{fig3} shows the result of molecular
dynamics simulations where we study the
energy loss of the molecule $E_{\mathrm{loss}}^{\rm  (M)}=E_{\rm{kin}}^{\rm (M)}-(1/2)m_{M}v_{M}^{\prime 2}$  for different mass ratios and laser parameters as in Fig.~\ref{fig2}. In the figure, the dashed and continuous lines correspond to pure sympathetic cooling and the predictions of Eqs.~\eqref{eq:vend}, respectively, while squares and dots are numerical results for different values of $E_{\rm{kin}}^{\rm (M)}$, averaged over 200 simulations. For $m_{A}<m_{M}$ the dominant energy loss mechanism is sympathetic cooling. However, for $m_{A}>m_{M}$ the energy loss of the molecule is mainly caused by dissipative collisional cooling, and is of the order of $\delta$, as expected. The {\it effective} atomic mass may be tuned using external confining potentials, e.g., optical lattices. For example, atoms can be confined in an optical trap with depth $V_{\rm tr} \gtrsim 2m_{A}V^{2}$, which for NaH with $v_{M}=\sqrt{3k_{B}T/m_{M}}$ and $T=0.5$ mK and Cs atoms with $m_{A}/m_{M}\approx 5.5$ implies $V_{\rm tr}>0.78 k_{B}T\approx 0.4$ mK.

\begin{figure}[tb]
\centering
\includegraphics[width=0.95\columnwidth]{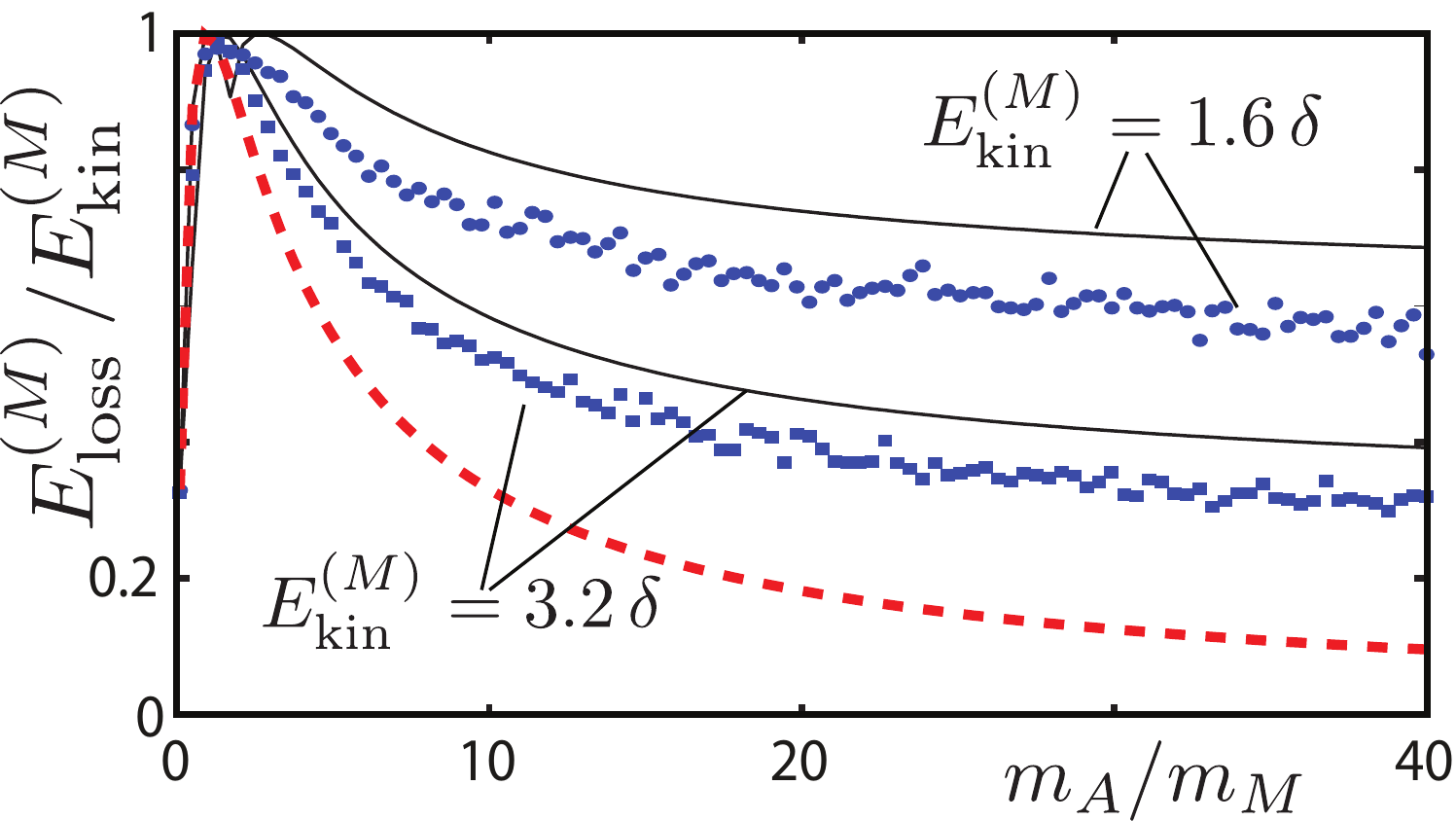}
\caption{Energy loss vs mass ratio $m_{A}/m_{M}$. The dashed red and solid black lines are analytic results for a model of sympathetic cooling only ($\delta = 0$) and finite $\delta$, respectively, see Eqs.~\eqref{eq:vend} and text. Blue dots and squares are averages over 200 runs of molecular dynamics simulations for finite $\delta$, for laser parameters as in Fig.~\ref{fig2}.}
\label{fig3}
\end{figure}

In conclusion, we have discussed a scenario where a molecule scatters successively from cold (stationary) atoms in designed elastic and inelastic processes. In this situation reminiscent of a ``microscopic version of a pinball machine'',  inelastic scattering events are associated with the emission of a photon implying a ``collisional Sisyphus'' cooling. While we focused on the simplest possible setup based on Van der Waals interactions, variants based on, e.g.~ dipole-dipole interactions and low dimensional trapping geometries seem possible. We will investigate the role of many-atom interactions in the dynamics of the gas in future work.

{\em Note added:} In the final stages of work we became aware of S.D.~Huber and H.P.~B\"uchler's proposal for Doppler cooling of polar molecules, where atomic Rydberg excitations serve as a bath for rotational molecular excitations~\cite{Buchler2011}.

We thank F.~Herrera for discussions, and S.D. Huber and H.P. B\"uchler for sharing with us their work before publication. Work supported by the Austrian Science Fund,  EU grants AQUTE, COHERENCE and NAME-QUAM, and by MURI, AFOSR, and EOARD.

\section*{Appendix}

\emph{Non-adiabatic transition}: We calculate the classical trajectory by $\mu \ddot{\mathbf{r}_{\mathrm{cl}}}%
=-\nabla V_{1}(\mathbf{r}_{\mathrm{cl}})$. Plugging the trajectory into the
Schr\"{o}dinger equations governing the dynamics of internal states and
calculating the transition probabilities after the molecule has reached the
flat region.\ The results are shown in Fig.~\ref{fig_s}, with laser parameters
as in Fig.~\ref{fig2}. $p_{1}$ is the computed transition probability from $%
|1(r)\rangle $ to the BO eigenstate that connect to $|r,0\rangle $ for $r\gg
r_{c}$, which is on the order of $10^{-2}$ for large kinetic energies $E_{%
\mathrm{kin}}\leq 3\delta .$ $p_{2}$ is the transition probability from $%
|1(r)\rangle $ to the BO eigenstate that connect to $|e,0\rangle $ for $r\gg
r_{c}^{\prime }$, which is tolerant as discussed. $p_{3}$ is
the non-adiabatic transition probability in $|2(r)\rangle $ at $%
r_{c}^{\prime \prime },$ which is calculated in a similar way. Note that
non-adiabatic transition probability in $|1(r)\rangle $ at $r_{c}^{\prime
\prime }$ is not important, since spontaneous decay almost takes place
deterministically. All the non-adiabatic transitions can be further suppressed
by increasing the Rabi frequency.

\begin{figure}[tb]
\centering
\includegraphics[width=\columnwidth]{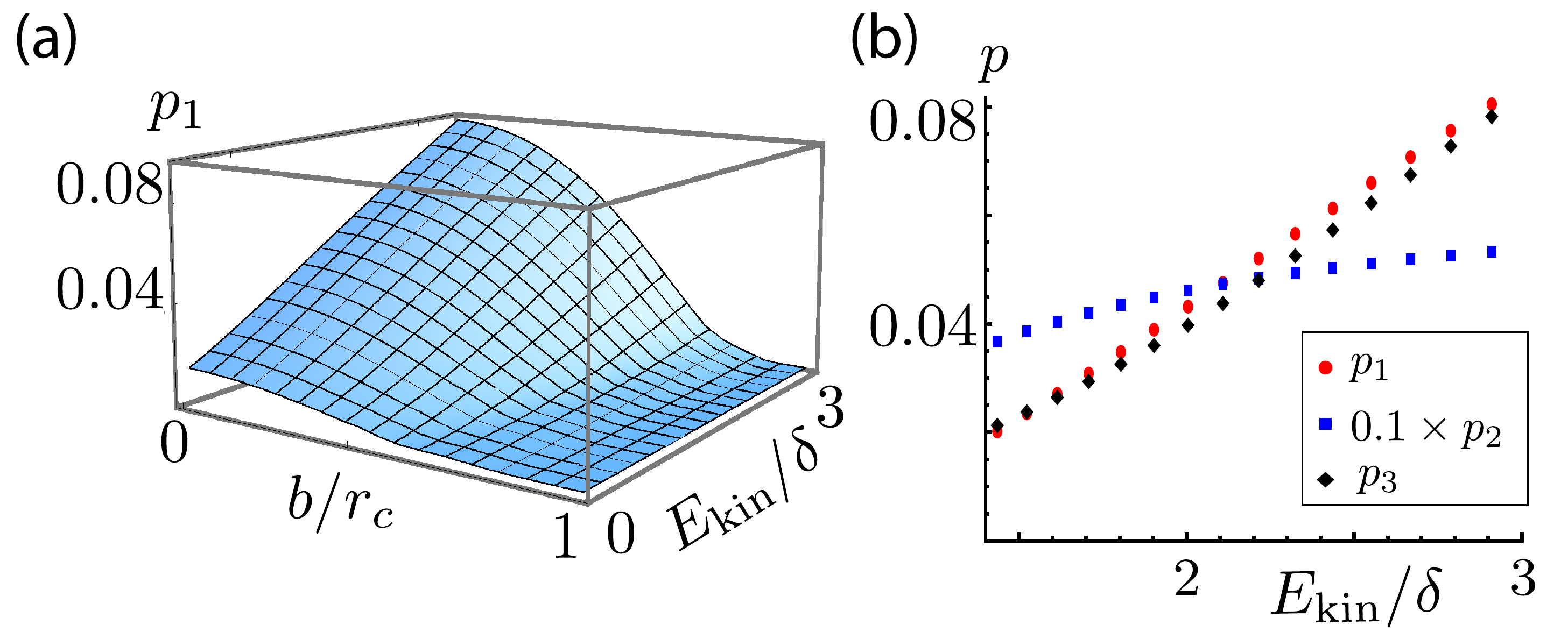}
\caption{Diabatic transitions between different BO eigenstates. (a) $p_{1}$ versus impact
parameter $b$ and initial kinetic energy. (b) Transition probabilities for $b=0$.}
\label{fig_s}
\end{figure}


\begin{thebibliography}{39}
\bibitem{comment0}For simplicity we do not explicitly reveal the degeneracy of the $p$-Rydberg-states in $|r'\rangle$.
\bibitem{comment1} Blue shielding at $r_{c}''$ can be achieved by coupling $|e\rangle$ or $|g_{2}\rangle$ to Rydberg states that have strong repulsive interaction with the molecule.
\bibitem{comment2} We may couple $|e\rangle$ ($|g_{2}\rangle$) to Rydberg states $|45s\rangle$ ($|44s\rangle$) with a blue detuning $0.92\Delta_{r}$ ($0.67\Delta_{r}$) and Rabi frequency $0.33\Delta_{r}$ ($0.37\Delta_{r}$) to engineer the blue shielding at $r_{c}''\approx 177$ nm.
\end{thebibliography}
\end{document}